\pdfoutput=1 
\documentclass{JINST}
\usepackage{subfigure}
\usepackage{amsmath}
\usepackage{upgreek}

\title{The Dynamic Range of LZ}

\author{J. Yin$^a$\thanks{Corresponding author.}~
for the LZ collaboration\\
\llap{$^a$}Department of Physics and Astronomy, University of Rochester,\\
  Rochester, New York 14627, USA\\
E-mail: \email{yin@pas.rochester.edu}}

\abstract{The electronics of the LZ experiment, the 7-tonne dark matter detector to be installed at the Sanford Underground Research Facility (SURF), is designed to permit studies of physics where the energies deposited range from 1~keV of nuclear-recoil energy up to 3,000~keV of electron-recoil energy.   The system is designed to provide a 70\% efficiency for events that produce three photoelectrons in the photomultiplier tubes (PMTs).  This corresponds approximately to the lowest energy threshold achievable in multi-tonne time-projection chambers, and drives the noise specifications for the front end.  The upper limit of the LZ dynamic range is defined to accommodate  the electroluminescence (S2) signals.  The low-energy channels of the LZ amplifiers provide the dynamic range required for the tritium and krypton calibrations. The high-energy channels provide the dynamic range required to measure the activated Xe lines. }


\begin{document}

\section{Introduction}\label{sec:intro}

LZ is a second-generation dark matter search experiment that will be installed at the Sanford Underground Research Facility (SURF) in South Dakota \cite{URLsurf}.  The design of LZ is described in detail in the LZ Conceptual Design Report \cite{LZcdr}.  The total mass of xenon in the time-projection chamber (TPC) is 7 tonne.  A liquid scintillator shield surrounds the cryostat.  Additional background reduction is achieved by instrumenting the xenon skin of the cryostat, that is the region between the field cage and the wall of the cryostat, with photomultiplier tubes (PMTs).  Simulations show that LZ should be able to detect (or exclude) Weakly Interactive Massive Particles (WIMPS) with a spin independent cross section of $2\times10^{-48}$~cm$^2$ per nucleon at a WIMP mass of 50~GeV/c$^2$ \cite{LZcdr}.

Events in the liquid xenon (LXe) target create excited and ionized atoms.  The excited atoms combine with un-excited atoms to form excitons which decay with the emission of ultraviolet photons (S1).  Electrons escaping recombination at the event site are caused to drift to the liquid surface and are extracted into the gas phase by applied electric fields, where they create electro-luminescence light (S2). Both signals are measured by arrays of PMTs, located above and below the active LXe region. Most of the S1 signal is measured with the bottom PMT array, as photons in the liquid are mostly trapped by total internal reflection at the liquid-gas interface and the polytetrafluoroethylene (PTFE) TPC lining.  The top PMT array images the $x-y$ location of the S2 signal and thus the $x-y$ location of the event site. The drift time, obtained from the time difference between the S1 and S2 signals, gives the $z$ position of the interaction. This technique thus provides 3D imaging of the event location.

In this paper, the dynamic range of LZ is discussed.  The LZ electronics was optimized to provide a 90\% efficiency to detect single photoelectrons and handle energy depositions associated with various calibration sources without saturating any component of the electronics chain.  Calibrations include electron-recoil calibrations using CH$_3$T (<~18.6~keV$_\textrm{ee}$), purity measurements using $^{83m}$Kr (41.6~keV$_\textrm{ee}$), nuclear-recoil calibrations using DD neutrons (<~74~keV$_\textrm{nr}$), and measurements of the $^{129m}$Xe activation line (236~keV$_\textrm{ee}$).  In addition, good energy and position reconstruction for larger energy depositions, such as those associated with $0\nu\beta\beta$ decay of $^{136}$Xe (2,458~keV$_\textrm{ee}$), is desired.  For large energy depositions, the large S2 signals will saturate a number of top PMTs.  For these events, the energy can still be determined using the bottom PMTs.  The impact on position reconstruction remains to be determined.

\section{Signal Processing Hardware}\label{sec:hardware}
\begin{figure}[t]
\centering
\includegraphics[width=0.75\textwidth]{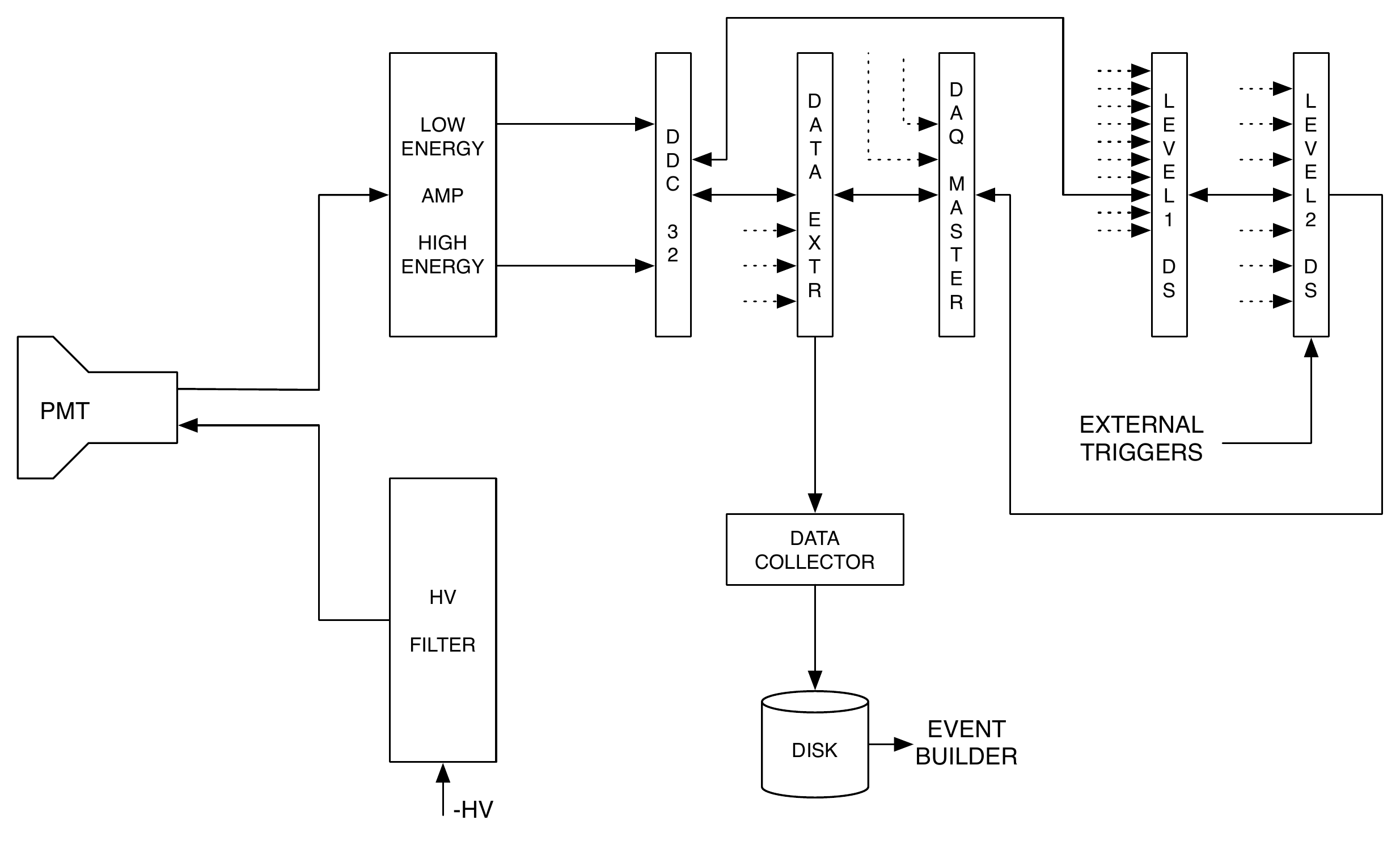}
\caption{A schematic of the signal processing of the TPC PMTs. The PMT signals are amplified and shaped using dual-gain amplifiers and the signals are digitized using DDC-32 digitizers.  }
\label{fig:SignalFlow}
\end{figure}

The signal processing of the TPC PMTs is shown schematically in Fig.~\ref{fig:SignalFlow}.  The PMTs operate at negative high voltages (HV), supplied by the LZ HV system. The PMT signals are processed by dual-gain amplifiers.  The amplified and shaped signals are connected to the data acquisition system (DAQ), described in detail in Chapter~11 of Ref.~\cite{LZcdr}. The digitized data are processed with Data Sparsifiers (DS) which make the decision whether or not to preserve the data.  The digitized data are extracted using Data Extractors (DE), sent to Data Collectors (DC), and stored on local disks. 

The dynamic range of LZ depends on the operating conditions of the TPC (e.g. the electric field in the TPC, the anode-gate voltage, and the pressure in the gas gap which influence the scintillation yield and number of ultraviolet photons produced for each ionization electron entering the gas phase), the linearity of the PMTs and the amplifiers, and the dynamic range of the digitizers.  In the remainder of this section, we will discuss the properties of these elements of the signal processing chain.

\subsection{Photomultiplier Tubes}\label{sec:pmts}

The LZ PMTs operate at a gain of $3.5\times10^6$.  The PMT base is designed to provide a linear response (non-linearities <~1\%) for S1 signals in all PMTs for energy depositions up to 2,458~keV.  Taking into consideration the 50~$\Omega$ load resistor in the base and the 32\% are loss in the signal cables, a single photoelectron will produce a 9.5~mVns signal at the input of the amplifier.

A 1~$\upmu$s square S2 pulse with up to 15,000 photoelectrons will not saturate the PMT.  Such pulse will produce a signal with an amplitude of less than 440~mV at the PMT output.  The S2 signals associated with a 2,458~keV energy deposition will saturate some of the PMTs on the top array but none of the PMTs on the bottom array, thus allowing the area of the S2 signals to be reconstructed accurately on the basis of the information provided by the bottom PMTs.  The saturation of up to 19 of the top PMTs, located above the interaction point, will reduce the accuracy of the $x-y$ position reconstruction of the interaction point.  The extent of the loss of accuracy remains to be determined.

\subsection{Amplifiers}\label{sec:amps}

The LZ amplifiers for the TPC PMTs have dual gains.  The low-energy output has an area gain of 40 and a shaping constant of  60~ns full-width tenth maximum (FWTM).  The high-energy output has an area gain of 4 and a shaping constant of  30~ns (FWTM).  The maximum amplitude at either output is 2.6~V.

\subsection{Digitizers}\label{sec:ddc}

The LZ digitizers sample the waveforms at 100~MHz with 14 bits.  The analog-to-digital converters (ADCs) have a dynamic range of 2~V.  Allowing for a 0.2~V undershoot, the effective dynamic range of the digitizers is 1.8~V.  

\section{Dynamic Range}\label{sec:DR}

The scintillation and ionization yields used to determine the dynamic range of LZ were obtained with the Noble Element Simulation Technique (NEST) \cite{Nest1}.  The yields for various calibration sources are summarized in Table~\ref{tab:NEST}.  The values listed in this table were obtained for a 700~V/cm electric field in the TPC.

\begin{table}[tbp]
\centering
\begin{tabular}{c c c c c}
\hline\hline
Source & Mode & Energy & Scintillation Yield & Ionization Yield \\
 & & (keV) & (S1, photons/keV) & (S2, electrons/keV) \\
\hline
CH$_3$T&ER&<~18.6&43&30\\
$^{83m}$Kr&ER&41.5&39&34\\
DD&NR&<~74&12&3\\
$^{129m}$Xe&ER&236&30&43\\
$0\nu\beta\beta$&ER&2,458&29&53\\
\hline
\end{tabular}
\caption{Scintillation and ionization yields in LZ for various sources.  The yields were obtained with the Noble Element Simulation Technique (NEST) \cite{Nest1} for a 700~V/cm electric field in the TPC.  For each source, the energy range is indicated as well as whether the interaction involves electron recoils (ER) or nuclear recoils (NR).}
\label{tab:NEST}
\end{table}

\subsection{S1 Dynamic Range}\label{sec:S1DR}

A single photoelectron in a PMT generates a signal at the digitizer with amplitudes of 87 ADC Counts (ADCC) and 17 ADCC for the low-energy and high-energy channel, respectively.  A 90\% efficiency for single photoelectrons in the low-energy channel requires a threshold of 33~ADCC.
Noise measurements have shown that the electronics noise in the system is less than 3~ADCC~(RMS).  The noise is thus more than 5$\sigma$ away from the threshold required to have a 90\% single photoelectron efficiency.

\begin{figure}
	\begin{minipage}[t]{.49\textwidth}
		\centering
		\includegraphics[width=1.0\linewidth]{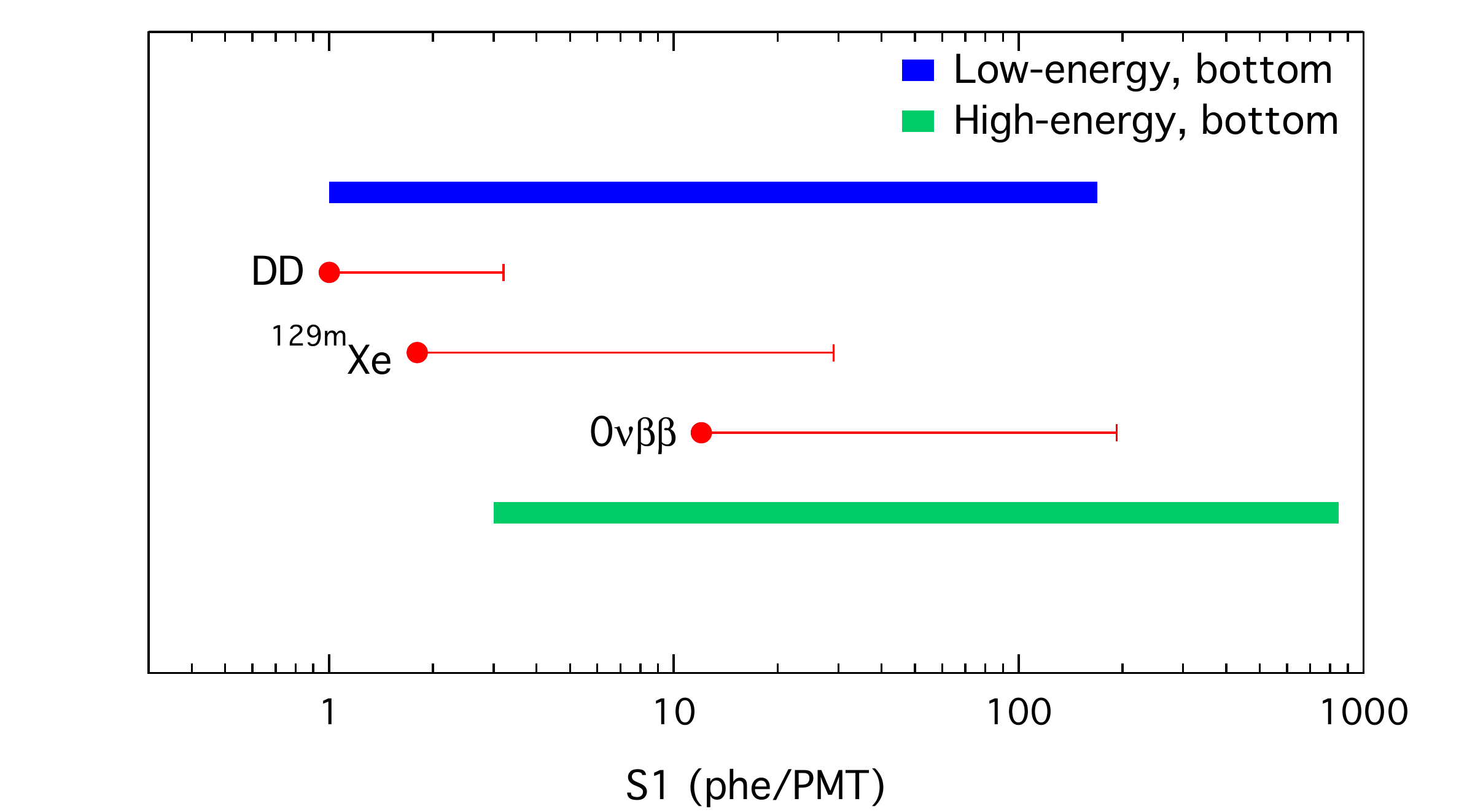}
		\caption{S1 dynamic range, expressed in terms of the number of photoelectrons per bottom pmt. The expected response for various sources is indicated.  The left-end of each source line corresponds to an energy deposition in the center of the TPC; the right-end corresponds to an energy deposition 1 cm above the cathode.}
		\label{fig:S1DR}
	\end{minipage}
	\hfill
	\begin{minipage}[t]{.49\textwidth}
		\centering
		\includegraphics[width=1.0\linewidth]{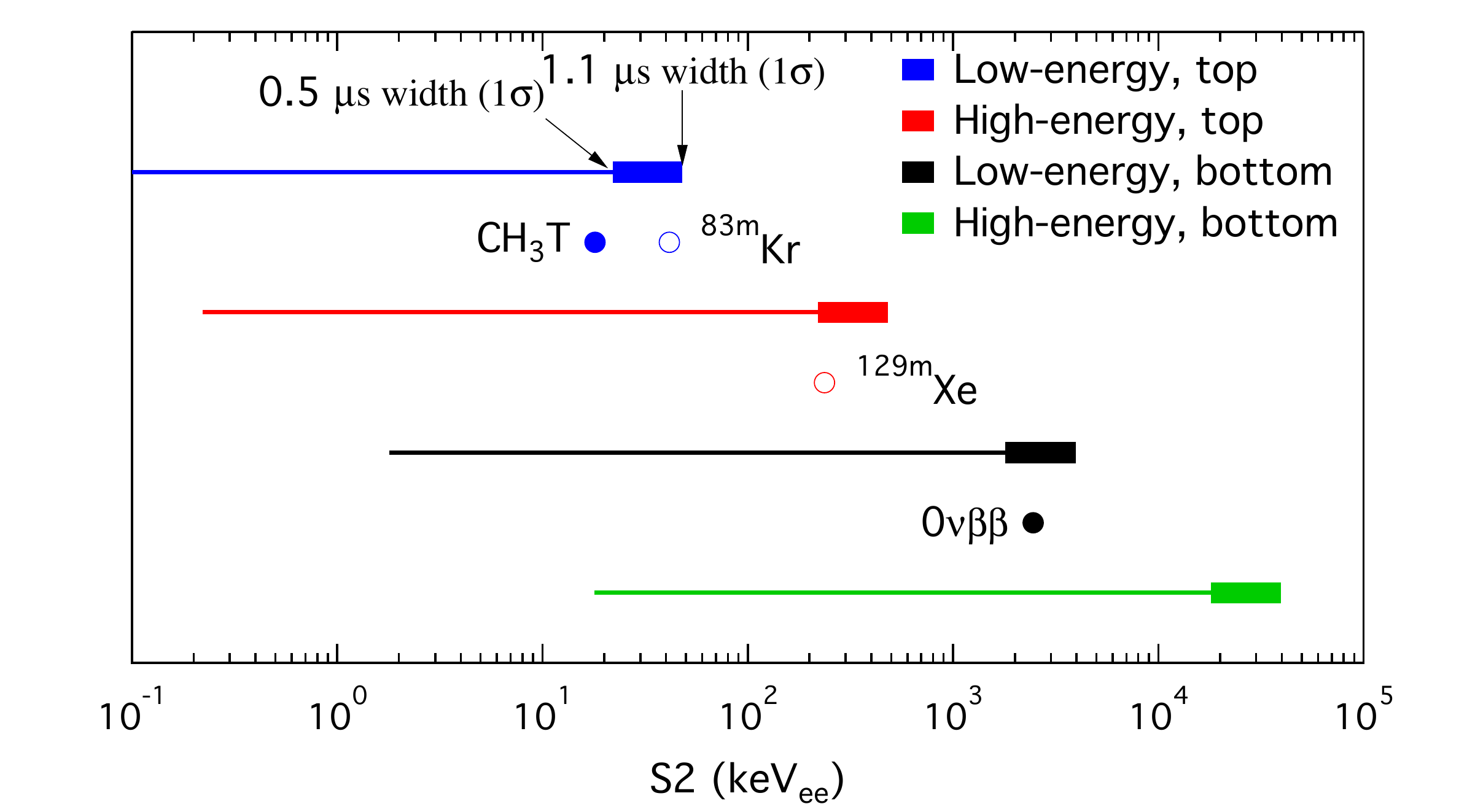}
			\caption{S2 dynamic range, expressed in terms of electron-recoil energy depositions (keV$_\textrm{ee}$) for the low- and high-energy channels of the top and bottom PMT arrays.  Energy depositions for various calibration sources are indicated by the circles.  The S2 signals associated with 74~keV$_\textrm{nr}$ from DD neutrons is smaller than the signals associated with the end point of CH$_3$.}
		\label{fig:S2DR}
	\end{minipage}
\end{figure}

The upper limit of the S1 dynamic range is defined by those signals with a 1.8~V amplitude at the input of the ADC.  The upper limits are 170~photoelectrons  and 850~photoelectrons for the low-energy and high-energy channels, respectively.  The lower and upper limits are indicated by the bars in Fig.~\ref{fig:S1DR}.

Simulations show that the photon detection efficiency, defined as the fraction of scintillation photons that produce photoelectrons in the LZ PMTs, depends on the position of the energy deposition.  Assuming a 95\% reflectivity of the PTFE used for the wall of the TPC, a 20\% reflectivity of the stainless steel used for the electrodes, and a 100~m absorption length of the photons in LXe, an average photon detection efficiency of 7.5\% is obtained \cite{LZcdr}.  
To calculate the maximum number of photoelectrons observed in a PMT, two scenarios are considered.  In the first scenario the photons are assumed to be emitted from the center of the detector.  In this scenario,  80\% of the S1 light is detected with the bottom array and the light distribution is uniform across the PMTs.  This corresponds to a photon detection efficiency of 0.025\% in a single bottom PMT.  In the second scenario, the photons are assumed to be emitted from a location 1~cm above the cathode.  Simulations show that in this scenario, the PMT right below the interaction point has a photon detection efficiency of 0.4\%.  The range of the S1 signal size for these two scenarios for several interactions of interest are shown in Fig.~\ref{fig:S1DR}.  S1 signals associated with DD neutrons and the $^{129m}$Xe activation line do not saturate the electronics; S1 signals associated with $0\nu\beta\beta$ decay of $^{136}$Xe (2,458~keV) can saturate the low-energy channels, but do not saturate the high-energy channels.

\subsection{S2 Dynamic Range}\label{sec:S2DR}

To calculate the dynamic range for S2 signals, we have assumed a 100\% extraction efficiency for the ionization electrons and that each ionization electron produces 500 ultraviolet photons in the gas phase.  The yield of ionization electrons for electron recoils varies between 27~e$^-$/keV at low energies (27~keV$_\textrm{ee}$) and 53~e$^-$/keV at high energies (2.5~MeV$_\textrm{ee}$).  For nuclear recoils, the yield of ionization electrons is considerably less; e.g. for DD neutrons the yield is 3~e$^-$/keV at 74~keV$_\textrm{nr}$.
The S2 light detected on the top PMT array is concentrated in a single PMT.  Simulations show that the photon detection efficiency in this PMT is 2\%.  The dynamic range for the S2 light detected with the top array is limited by the linearity of the PMTs and the dynamic range of the digitizers.  The PMTs are not saturated by 1~$\upmu$s square S2 pulses with an area up to 15,000~photoelectrons; such pulses produce signals of up to 440~mV at the PMT output.  Taking into consideration the attenuation in the signal cables, these signals will produce 1.3~V pulses from the high-energy channels at the digitizer.  Non-linear effects in the PMTs become important at the time the high-energy signals reach amplitudes similar to the 1.8~V upper limit of the digitizer.  

The upper limits of the dynamic range for the low-energy and high-energy channels of the top PMTs is shown by the top two bars in Fig.~\ref{fig:S2DR}.  The dynamic range depends on the width of the S2 signal and two limits are shown.  The left-hand and right-hand sides of each bar show the upper limits of the dynamic range for 0.5~$\upmu$s (1$\sigma$) and 1.1~$\upmu$s (1$\sigma$) wide S2 signals, respectively.  The energy depositions associated with CH$_3$T,  $^{83m}$Kr, and $^{129m}$Xe  decays are well below the upper limit of the S2 dynamic range.  Measurements of the energy depositions associated with $0\nu\beta\beta$ decay of $^{136}$Xe will saturate one or more PMTs of the top array.

The S2 light detected on bottom PMT array is distributed rather uniformly across the array.  Simulations show that the maximum photon detection efficiency in a bottom PMT is 0.012\%.  Since the fraction of S2 light that ends up on the bottom array is independent of S2 signal size, the yield observed with just the bottom PMT array can be used to reconstruct the size of the S2 signal.  The upper limits of the dynamic range for S2 signal detected with the bottom array is shown by the bottom two bars in Fig.~\ref{fig:S2DR}.  These calculations show that the S2 signal size can be reconstructed for every depositions well above the energy deposition associated with $0\nu\beta\beta$ decay of $^{136}$Xe.

\section{Summary and Conclusions}\label{sec:summary}
The design of the PMT base and the analog and digital electronics has been optimized to maximize the dynamic range of LZ.  The dynamic range exceeds the dynamic range required for normal dark-matter search mode (<~6~keV$_\textrm{ee}$ and <~30~keV$_\textrm{nr}$ \cite{LUXprl}) and the regular detector calibrations.   Energy deposition of several MeV can be observed without saturation using the bottom PMT array.

\acknowledgments

The work was partially supported by the U.S. Department of Energy (DOE) under award numbers DE- SC0012704, DE-SC0010010, DE-AC02-05CH11231, DE-SC0012161, DE-SC0014223, DE-FG02- 13ER42020, DE-FG02-91ER40674, DE-NA0000979, DE-SC0011702, DE-SC0006572, DE-SC0012034, DE-SC0006605, and DE-FG02-10ER46709; by the U.S. National Science Foundation (NSF) under award numbers NSF PHY-110447, NSF PHY-1506068, NSF PHY-1312561, and NSF PHY-1406943; by the U.K. Science \& Technology Facilities Council under award numbers ST/K006428/1, ST/M003655/1, ST/M003981/1, ST/M003744/1, ST/M003639/1, ST/M003604/1, and ST/M003469/1; and by the Portuguese Foundation for Science and Technology (FCT) under award numbers CERN/FP/123610/2011 and PTDC/FIS-NUC/1525/2014.

\end{document}